\documentclass[a4paper,11pt]{article}
\usepackage{pos}

\usepackage{float}
\newcommand{\GeV}{\mathrm{GeV}}

\title{
{\footnotesize CERN-TH-2024-100, ZU-TH 31/24, RISC Report number 24-04,  DESY-24-096, PoS (LL2024) 047}\\
The three-loop single-mass heavy flavor corrections to 
deep-inelastic 
scattering} 
\ShortTitle{The three-loop heavy flavor corrections}

\author[a,b]{J. Ablinger}
\author[c]{A. Behring}
\author*[d,e]{J. Bl\"umlein}
\author[a,d]{A. De Freitas}
\author[f]{A. von Manteuffel}
\author[a]{\newline C. Schneider} 
\author[g]{K. Sch\"onwald}

\affiliation[a]{Johannes Kepler University,
Research Institute for Symbolic Computation (RISC), \newline
Altenberger Stra\ss{}e 69, A-4040, Linz, Austria}

\affiliation[b]{Johann Radon Institute for Computational and Applied Mathematics
(RICAM), Austrian Academy of Sciences, Altenberger Stra\ss{}e 69, A-4040 Linz Austria}

\affiliation[c]{Theoretical Physics Department, CERN, 1211 Geneva 23, Switzerland}

\affiliation[d]{Deutsches Elektronen-Synchrotron DESY, Platanenallee 6, 15738 Zeuthen, Germany}

\affiliation[e]{Institut f\"ur Theoretische Physik III, IV, TU Dortmund, Otto-Hahn Stra\ss{}e 4, \newline 44227
Dortmund, Germany}

\affiliation[f]{Institut f\"ur Theoretische Physik, Universit\"at Regensburg,
93040 Regensburg, Germany}

\affiliation[g]{Physik-Institut, Universit\"at Z\"urich,
Winterthurerstrasse 190, CH-8057 Z\"urich, Switzerland}

\emailAdd{Johannes.Bluemlein@desy.de}

\abstract{
We report on the status of the calculation of the massive Wilson coefficients and operator matrix elements 
for deep-inelastic scatterung to three-loop order. We discuss both the unpolarized and the polarized case, 
for which all the single-mass and nearly all two-mass contributions have been calculated. Numerical 
results on the structure function $F_2(x,Q^2)$ are presented. In the polarized case, we work in the Larin
scheme and refer to parton distribution functions in this scheme. Furthermore, results on the three-loop 
variable flavor number scheme are presented.}

\FullConference{Loops and Legs in Quantum Field Theory (LL2024)\\
 14-19, April, 2024\\
Wittenberg, Germany\\}

\begin{document}
\maketitle

%---------------------------------------------------------------------------------
\section{Introduction}
\label{sec:1}
%---------------------------------------------------------------------------------

\vspace*{1mm}
\noindent
The massive operator matrix elements (OMEs) \cite{Buza:1995ie,Ablinger:2017err} and massive Wilson 
coefficient of deep-inelastic scattering \cite{Buras:1979yt,Reya:1979zk,Blumlein:2012bf,Blumlein:2023aso} 
in the asymptotic region $Q^2 \gg m^2$, with $m$ the heavy quark mass, have been calculated in the single-mass 
case to three-loop order in Quantum Chromodynamics (QCD). To two-loop order 
these OMEs have been computed 
in Refs.~\cite{Witten:1975bh,Watson:1981ce,Buza:1995ie,Buza:1996wv,Buza:1996xr,
Bierenbaum:2007qe,Bierenbaum:2007pn,Bierenbaum:2008yu,Bierenbaum:2009zt,Behring:2014eya,Blumlein:2016xcy,
Blumlein:2019qze,Blumlein:2019zux,Blumlein:2021xlc,Bierenbaum:2022biv}. 
At three-loop order seven unpolarized massive OMEs, $A_{qq,Q}^{\rm NS}, A_{Qq}^{\rm PS}, 
A_{qq,Q}^{\rm PS}, A_{Qg}, A_{qg,Q}, A_{gq,Q}$ and $A_{gg,Q}$, and the corresponding polarized OMEs 
contribute. First a series of Mellin moments was calculated in Ref.~\cite{Bierenbaum:2009mv}. 
The computation of theses
functions for general values of Mellin-$N$ followed in Refs.~\cite{Ablinger:2010ty,Behring:2014eya,
Ablinger:2014lka,Ablinger:2014vwa,Ablinger:2014nga,Behring:2015zaa,Blumlein:2016xcy,
Ablinger:2019etw,Behring:2021asx,Blumlein:2021xlc,Ablinger:2022wbb,Ablinger:2023ahe,Ablinger:2024xtt}. 
Two-mass corrections contribute starting from two-loop order, i.e. at next-to-leading-order (NLO),
cf.~\cite{Blumlein:2018jfm}, as factorizable terms.
From three-loop order onward also irreducible two-mass terms contribute, cf. 
Refs.~\cite{Ablinger:2017err,Ablinger:2017xml,Ablinger:2018brx,Ablinger:2019gpu,Ablinger:2020snj}. 
The last missing term of this class will be published soon 
\cite{TWOMAQG}. Also, for charged current 
structure functions a series of 
heavy-flavor corrections was calculated \cite{Buza:1997mg,Blumlein:2014fqa,
Behring:2015roa,Behring:2016hpa}. The massive Wilson coefficients  depend on the massless three-loop
unpolarized Wilson coefficients \cite{Vermaseren:2005qc,Blumlein:2022gpp} and the polarized ones
\cite{Blumlein:2022gpp}. The evolution of the massless parton densities to three-loop order
depend on the unpolarized 
\cite{Moch:2004pa,Vogt:2004mw,Ablinger:2014nga,Anastasiou:2015vya,Ablinger:2017tan,
Mistlberger:2018etf,Luo:2019szz, Duhr:2020seh,Ebert:2020yqt,Ebert:2020unb,Luo:2020epw,Blumlein:2021enk,
Blumlein:2022gpp,Baranowski:2022vcn,Gehrmann:2023ksf}
and polarized \cite{Moch:2014sna,Behring:2019tus,Blumlein:2021ryt} three-loop anomalous dimensions.

The technical aspects of the calculation of these massive OMEs consist of a series of standard steps, 
described e.g. in
Ref.~\cite{Ablinger:2023ahe}. The integration-by-parts reduction has been performed using {\tt Reduze 2} 
\cite{Studerus:2009ye,vonManteuffel:2012np}. We used also more special analytic methods, such as summation 
and guessing 
methods applied to a very large number of moments \cite{SIG1,SIG2,SIG3,Blumlein:2017dxp,GUESS,Blumlein:2009tj,
ABRAMOV,CS1}, 
special higher transcendental function treatment of different kind 
\cite{Vermaseren:1998uu,
Blumlein:1998if,
Ablinger:2013cf,
Ablinger:2011te,
Ablinger:2014bra,
Remiddi:1999ew,
Blumlein:2003gb,
Blumlein:2009ta,
Blumlein:2009fz,
Ablinger:2010kw,
Ablinger:2013hcp,
Ablinger:2014rba,
Ablinger:2015gdg,
ALL2016,
ALL2018,
Ablinger:2018cja,
Ablinger:2019mkx,
Ablinger:2021fnc},
differential equation methods for first-order-factorizing systems \cite{Ablinger:2018zwz} and 
non-first-order-factorizing systems \cite{Ablinger:2017bjx,Behring:2023rlq}, 
including (general) semi-analytic solutions
\cite{Fael:2021kyg,Fael:2022miw}. In our calculations the use of the packages
{\tt Sigma} \cite{SIG1,SIG2,SIG3},
{\tt HarmonicSums} \cite{Vermaseren:1998uu,
Blumlein:1998if,
Ablinger:2013cf,
Ablinger:2011te,
Ablinger:2014bra,
Remiddi:1999ew,
Blumlein:2003gb,
Blumlein:2009ta,
Blumlein:2009fz,
Ablinger:2010kw,
Ablinger:2013hcp,
Ablinger:2014rba,
Ablinger:2015gdg,
ALL2016,
ALL2018,
Ablinger:2018cja,
Ablinger:2019mkx,
Ablinger:2021fnc},
{\tt OreSys}  \cite{ORESYS1,ORESYS2,ORESYS3}, and others \cite{Klappert:2019emp,
Klappert:2020aqs} played an important role.

The present note is organized as follows. In Section~\ref{sec:2} we present the three-loop single-mass 
contributions at next-to-next-to-leading order (NNLO) to the structure function $F_2(x,Q^2)$ for the first 
time. It is an important ingredient for precision QCD fits of the deep-inelastic World data to determine
the strong coupling constant $a_s = \alpha_s/(4\pi)$, cf.~\cite{Bethke:2011tr,Moch:2014tta,Alekhin:2016evh,
dEnterria:2022hzv} and of the parton distribution functions (PDFs), \cite{Accardi:2016ndt}.
To obtain the same results for the polarized structure functions one needs also PDFs evolved in the 
Larin scheme. This is discussed in Section~\ref{sec:3}. The three-loop massive OMEs also allow one to
derive the matching relations in the variable flavor number scheme (VFNS) at three-loop order,
which is presented in Section~\ref{sec:4}. Section~\ref{sec:5} contains the conclusions.
%---------------------------------------------------------------------------------
\section{\boldmath The Single-Mass Heavy-Flavor Contributions to $F_2(x,Q^2)$}
\label{sec:2}
%---------------------------------------------------------------------------------

\vspace*{1mm}
\noindent
The current results on the three--loop massive OMEs allow us to compute the structure
function $F_2(x,Q^2)$, including the massless and single-mass heavy-flavor corrections
due to $c$- and $b$-quarks at large enough scales $Q^2$.
The massive OMEs and massive asymptotic Wilson coefficients are calculated for quark masses in the
on-shell scheme, $m_c = 1.59~\GeV$, \cite{Alekhin:2012vu}, and $m_b = 4.78~\GeV$ \cite{Agashe:2014kda}.
It has been shown in Ref.~\cite{Buza:1995ie} that the criterion $Q^2 \gg m^2$, for which the asymptotic 
structure function represents the full  structure function
$F_2(x,Q^2)$ at the $1\%$-level is fulfilled by $Q^2/m^2 \geq 10$, i.e. for $Q^2 \geq 25~\GeV^2$ in the 
case of 
charm at NLO.\footnote{This criterion may be different in the case of other structure functions.} 

%---------------------------------------------------------------------------------
\begin{figure}[H] \centering
\includegraphics[width=0.44\textwidth]{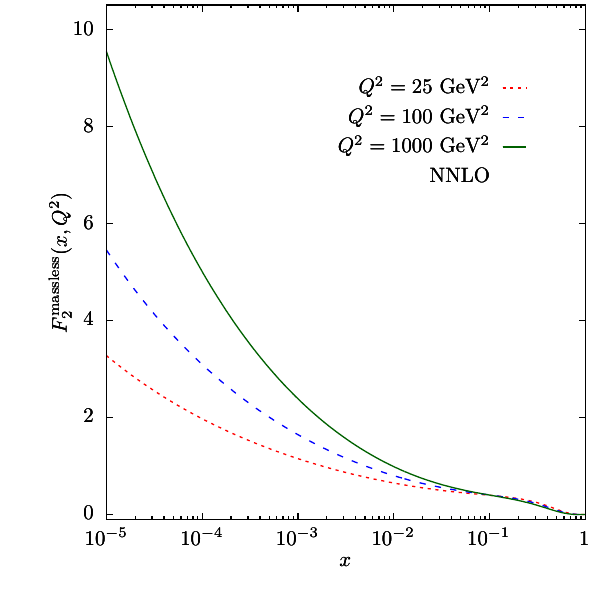}
\includegraphics[width=0.44\textwidth]{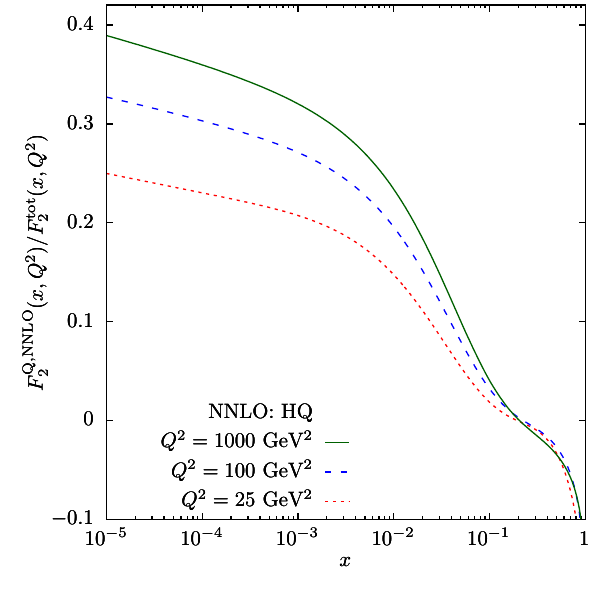}
\caption[]{Left panel: the massless contributions to the structure function $F_2(x,Q^2)$ at NNLO using the 
PDFs of Ref.~\cite{Alekhin:2017kpj}. 
Right panel: The ratio of the NNLO single-mass charm and bottom contributions to $F_2(x,Q^2)$ to its
total value. 
Dotted lines: $Q^2 =   25~\GeV^2$;
dashed lines: $Q^2 =   100~\GeV^2$;
full lines: $Q^2 = 10000~\GeV^2$.
\label{FIG1}}
\end{figure}
%---------------------------------------------------------------------------------

In Figure~\ref{FIG1} we present both the prediction for the massless contributions to the structure 
function $F_2(x,Q^2)$ as well as for  the single-mass $c$ and $b$-quark contributions at NNLO for a wide 
range in the kinematic variables Bjorken $x$ and the virtuality $Q^2$. The fraction of the (virtual and 
real) heavy quark contributions vary from $\sim 25 \%$ to $40 \%$ at $x = 10^{-4}$ for $Q^2$ in the 
range between $25~\GeV^2$ and $10^4~\GeV^2$ and the contribution falls towards large values of $x$.
Here five different massive Wilson coefficients contribute.

Already in 1990 the massive OME $A_{Qg}$ has been investigated in its ultimate small $x$ limit to any 
order in $a_s$, using methods of $k_\perp$-factorization \cite{Catani:1990eg}. In 1995 the respective
expansion term of $O(a_s^2)$ has been confirmed by expanding the complete NLO result in 
Ref.~\cite{Buza:1995ie}. After 34 years we have now confirmed also the $O(a_s^3)$
term for the first time in Refs.~\cite{Ablinger:2023ahe,Ablinger:2024xtt}. One obtains
%--------------------------------------------------------------------------------------------------------
\begin{equation}
a_{Qg}^{(3), x \rightarrow 0}(x) = \frac{64}{243} C_A^2 T_F \left[1312 + 135 \zeta_2 - 189 \zeta_3 \right] 
\frac{\ln(x)}{x}
\label{eq:SX}
\end{equation}
%--------------------------------------------------------------------------------------------------------
for the constant part of the unrenormalized three-loop OME $A_{Qg}^{(3)}$. Here
$C_A = N_c, T_F = 1/2, C_F = (N_c^2-1)/(2 N_c)$ denote the color factors, 
with $N_c = 3$ for QCD, and $\zeta_k$ are the values of
Riemann's $\zeta$-function at integer argument, $k \geq 2$.
However, this term does not describe the small $x$ behaviour, neither of the massive OME 
nor of the structure function, due
to very large sub-leading small $x$ corrections, as the numerical 
analysis in Ref.~\cite{Ablinger:2024xtt} shows. This is a quite common observation for a long list of BFKL
predictions.\footnote{For a survey, see Ref.~\cite{Blumlein:1999ev}.} Already in 
Ref.~\cite{Ablinger:2014nga} we have computed the pure-singlet OME $A_{Qq}^{(3),PS}$ and derived the
corresponding quantity $a_{Qq}^{(3),PS, x \rightarrow 0}(x)$ and its leading small $x$ limit. It is
related to (\ref{eq:SX}) through rescaling by the factor $C_F/C_A$, as has  been found by an explicit
analytic calculation now.
%---------------------------------------------------------------------------------
\section{Polarized Parton Distributions in the Larin Scheme}
\label{sec:3}
%---------------------------------------------------------------------------------

\vspace*{1mm}
\noindent
The three-loop massless \cite{Blumlein:2022gpp} and massive Wilson coefficients in the polarized
case were calculated in the Larin scheme \cite{Larin:1993tq,Matiounine:1998re}. Currently it is 
not possible to construct the transformation into the $\overline{\rm MS}$ scheme at three-loop order
for them. 
However, the polarized structure function $g_1(x,Q^2)$, as an observable, can be expressed in terms of 
the Wilson coefficients and parton distributions \cite{Blumlein:2024euz} in the Larin scheme. The 
scaling violations of the polarized massless parton densities are described in this scheme as well 
by using the corresponding anomalous dimensions \cite{Moch:2014sna,Blumlein:2021ryt}. In 
Ref.~\cite{Blumlein:2024euz} the polarized parton distribution functions have been provided in the 
Larin scheme up to NNLO recently.

\begin{figure}[H] \centering
\includegraphics[width=0.44\textwidth]{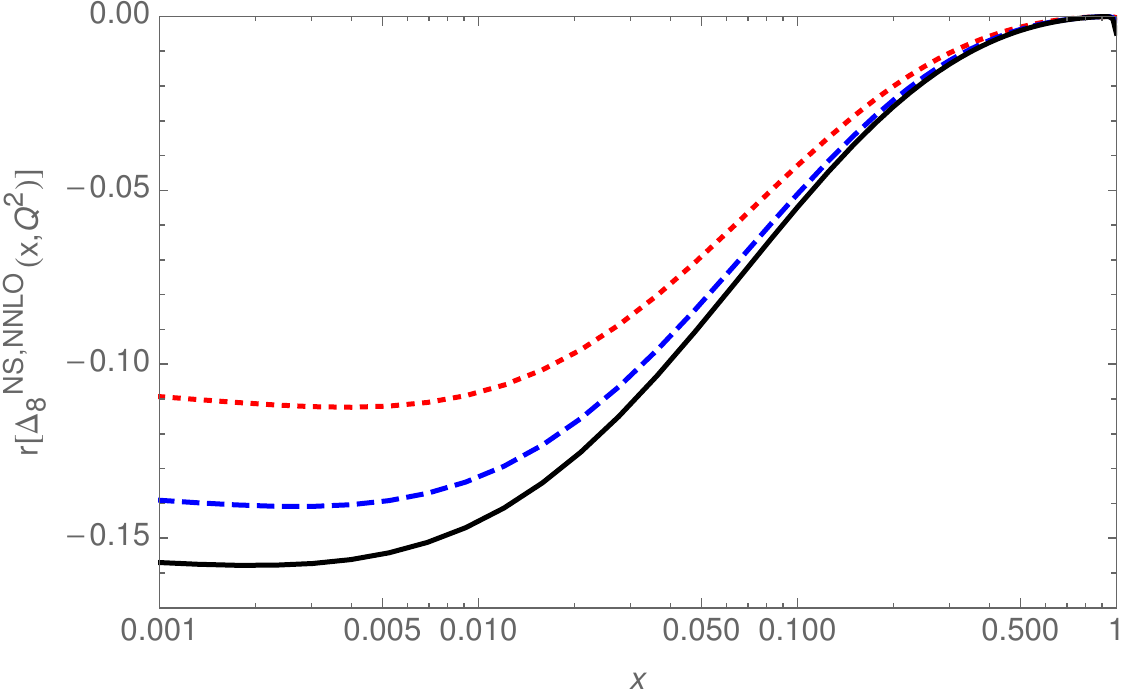}
\includegraphics[width=0.44\textwidth]{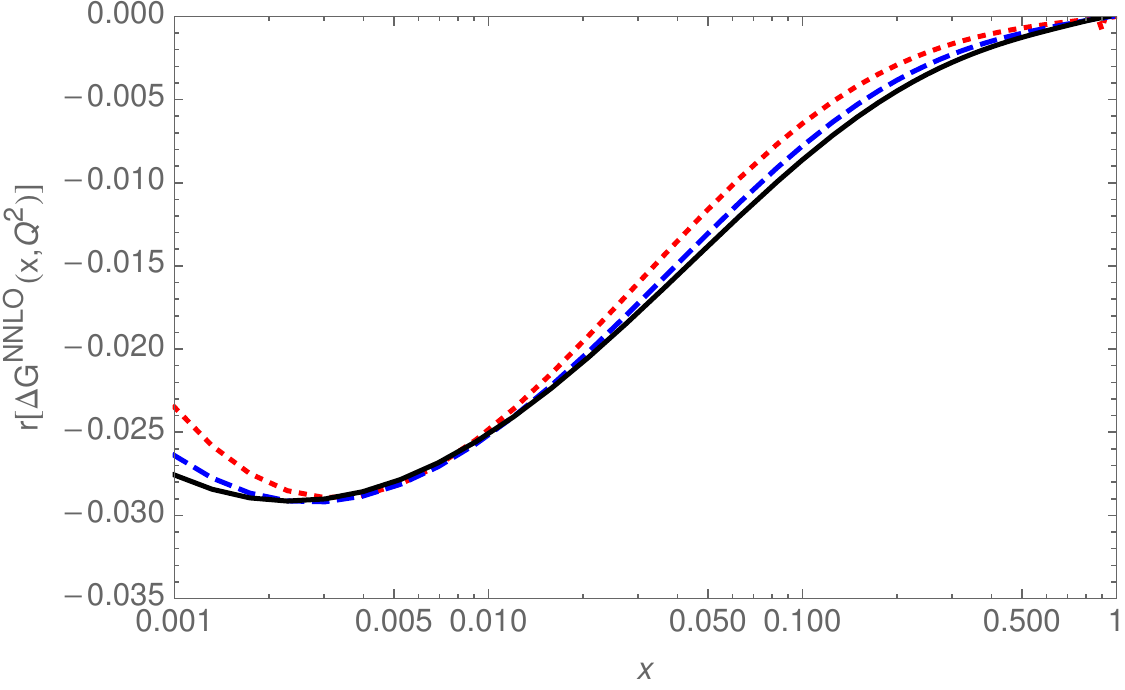}
\caption{The relative change of the polarized parton distribution functions 
$\Delta_8(x,Q^2) = \Delta u(x,Q^2) + \Delta d(x,Q^2)$ and 
$\Delta G(x,Q^2)$ comparing the evolution in the  $\overline{\rm MS}$ scheme and the Larin scheme.
Dotted line: $Q^2 = 100~\GeV$; dashed line: $Q^2 = 1000~\GeV$; full line: $Q^2 = 10000~\GeV$; from 
Ref.~\cite{Blumlein:2024euz}.
\label{FIG2}}
\end{figure}

Starting at NLO the scale evolution is different in the Larin and the $\overline{\rm MS}$ scheme. We 
illustrate this in Figure~\ref{FIG2} for the ratio $r = f^{\rm Larin}/f^{\overline{\rm MS}}-1$
at NNLO. The 
effects are larger for the quarkonic distributions than for 
the gluon distribution. In the latter case they are caused by mixing effects with the quarkonic anomalous 
dimensions only, since $\Delta P_{gg}^{(1,2)}$ is the same in both schemes. In the large $x$ limit the anomalous 
dimensions in both schemes approach each other.

\noindent
For the quark distributions the relative change in the small $x$ region, $x \sim 0.001$, 
may reach 10--15\%, while for the gluon distribution the corresponding effect amounts to 
$O(3\%)$. High precision QCD fits in the polarized case therefore require to use Larin-scheme 
PDFs. 
%---------------------------------------------------------------------------------
\section{The Single-Mass Variable Flavor Number Scheme}
\label{sec:4}
%---------------------------------------------------------------------------------

\vspace*{1mm}
\noindent
The single-mass three-loop massive OMEs allow one to construct the corresponding matching relations
in the VFNS to three-loop order. The principal structure of the matching relations has been given in 
Ref.~\cite{Buza:1996wv} and was corrected in Ref.~\cite{Bierenbaum:2009mv}. The VFNS relates the massless 
parton
densities with $N_F$ massless quark flavors to the ones of $N_F + 1$ massless quark flavors in the
region $Q^2 \gg m^2_Q$, where $m_Q$ is the mass of the heavy quark becoming effectively massless. In course
of this, one also obtains massive quark distributions, $f_Q(x,Q^2) + f_{\bar{Q}}(x,Q^2)$. The
relations are derived from the structure functions at high scales $Q^2$ in the fixed flavor number 
scheme and are determined by the process-independent massive OMEs. It has been shown in 
Refs.~\cite{Buza:1995ie,Buza:1996xr,Blumlein:2016xcy,Blumlein:2019qze,Blumlein:2019zux}
at two-loop order for the cases in which the complete heavy-quark-mass dependence is known analytically 
that 
the effective massless approach in the case of charm and bottom quarks  only applies at 
high scales $Q^2$ and neither at the scales $m_c^2$ nor $m_b^2$. The new PDFs obtained in the VFNS
are 
inserted
into the massless representation of the corresponding structure functions. If one analytically
expands the resulting expressions in the coupling constant $a_s$ one obtains
again the structure functions in the fixed flavor number scheme to the order one worked in.
The differences in the VFNS to the result in the direct calculation are of higher order in $a_s$.
Depending on the matching scale chosen, different size pile-up effects due to these terms
are obtained.

\begin{figure}[H] \centering
\includegraphics[width=0.44\textwidth]{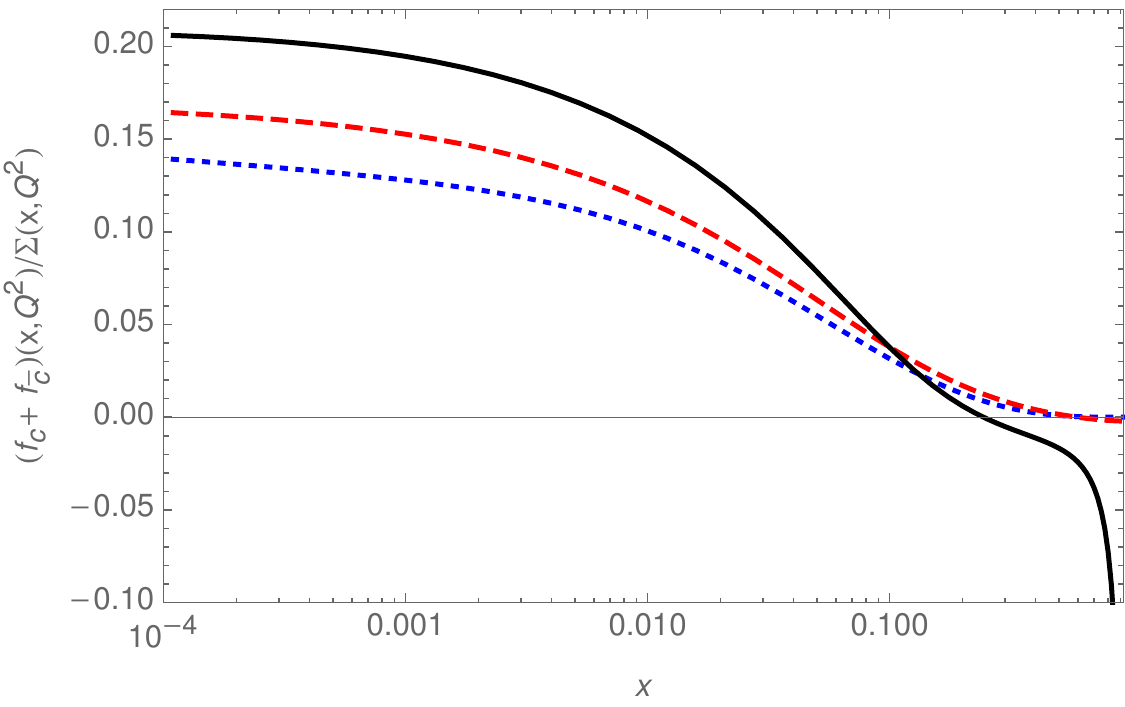}
\caption[]{The distribution $f_c(x,Q^2) + f_{\bar{c}}(x,Q^2)$ normalized to $\Sigma^{\rm NF=3}(x,Q^2)$.
Dotted lines: $Q^2 = 30~\GeV^2$; 
dashed lines: $Q^2 = 100~\GeV^2$; 
full lines: $Q^2 = 10000~\GeV^2$; from Ref.~\cite{VFNS}.
\label{FIG3}
}
\end{figure}

An implementation of the three-loop matching relations will be given in Ref.~\cite{VFNS}. 
In Figure~\ref{FIG3} we illustrate the charm distribution $f_c(x,Q^2) + f_{\bar{c}}(x,Q^2)$ normalized to the singlet
distribution for $N_F = 3$. The effects grow with $Q^2$ due to the logarithmic terms $\ln(m_Q^2/Q^2)$
in the matching relations. 
%---------------------------------------------------------------------------------
%---------------------------------------------------------------------------------

%---------------------------------------------------------------------------------
\section{Conclusions}
\label{sec:5}
%---------------------------------------------------------------------------------

\vspace*{1mm}
\noindent
We finished the calculation of all single-mass OMEs and asymptotic inclusive heavy-flavor
Wilson coefficients to three-loop order and made numerical predictions for the structure function
$F_2(x,Q^2)$. Very soon, also the two-mass corrections will also be finished both in the unpolarized and 
polarized cases. In the polarized case we worked in the Larin scheme and provided a first set of NNLO
parton densities. Furthermore, the single-mass matching relations in the VFNS are
now available both in the unpolarized and polarized cases. 

The present results are of special importance for phenomenological predictions of the precision physics
at future facilities such as the EIC \cite{Boer:2011fh} and LHeC \cite{LHeCStudyGroup:2012zhm,LHeC:2020van}, 
but also for re-analysis of the HERA \cite{Blumlein:1989pd} and other World deep-inelastic 
data, as well as for inclusive measurements at the LHC in its  high luminosity phase at CERN, and its future 
successor, the FCC \cite{FCC}. In the flavor non-singlet case, the relations for a QCD-fit in the unpolarized
and polarized cases were given in \cite{Blumlein:2021lmf} in the scheme-invariant representation, which allows
a direct fit of $a_s(M_Z)$ using measured input distributions at the starting scale $Q_0^2$.

\vspace*{2mm}
\noindent
{\bf Acknowledgment.} 
We would like to thank S.~Klein for calculating a series of Mellin moments in the polarized case
\cite{SKLEIN} by using {\tt MATAD} \cite{Steinhauser:2000ry}, which we have used for comparison, and we thank 
P.~Marquard for discussions. 
This work has been supported in part  by the Austrian Science Fund (FWF) 10.55776/P33530
and P34501N.

\end{document}